\documentclass[a4,11pt]{article}
\pdfoutput=1

\usepackage{graphicx}	
\usepackage{dcolumn}	
\usepackage{bm}		

\usepackage{rotating}

\mathchardef\nabla="7272

\begin{document}

\title{Direct Numerical Simulation of a separated channel flow \\ with a smooth profile}

\author{M. Marquillie, J.-P. Laval, R. Dolganov \\ \mbox{ }\\ Laboratoire de M\'ecanique de Lille, \\ CNRS UMR 8107, Blv. Paul Langevin,\\ F-59655 Villeneuve d'Ascq, France}

\maketitle

\begin{abstract}
A direct numerical simulation (DNS) of a channel flow with one curved surface was performed at moderate Reynolds number ($Re_\tau$ = 395 at the inlet). The adverse pressure gradient was obtained by a wall curvature through a mathematical mapping from physical coordinates to Cartesian ones. The code, using spectral  spanwise and normal discretization, combines the advantage of a good accuracy with a fast integration procedure compared to standard numerical procedures for complex geometries.  The turbulent flow slightly separates on the profile at the lower curved wall and is at the onset of separation at the opposite flat wall. The thin separation bubble is characterized with a reversal flow fraction. Intense vortices are generated near the separation line on the lower wall but also at the upper wall. Turbulent normal stresses and kinetic energy budget are investigated along the channel.\\


\end{abstract}

\section{Introduction}

In the last two decades, the control of flow around an airfoil is of strong interest. The aim is to delay or to suppress the detachment of the turbulent boundary layer in several flight phases of an airplane. The turbulent boundary layer submitted to an adverse pressure gradient is thus one of the main research effort in aeronautics and more generally in transport. It is imperative to improve our knowledge on near wall turbulence submitted to pressure gradient in order to develop new strategies for the control acting on the turbulent coherent structures.

Most of the knowledge on this type of flow have been obtained experimentally using hot wire anemometry. There is several ways to generate a pressure gradient experimentally, which lead to very different pressure gradient configurations, constant or varying in space, with or without separation. It is not the aim, here, to make an extended review of the subject but just to present relevant contributions related to the present study. Skare \& Krogstad~\cite{skare94,krogstad95} did a detailed study of a boundary-layer in a strong adverse pressure gradient with a constant skin friction coefficient, they provided a thorough analysis of the turbulence statistics of the flow, in particular the budget of the kinetic energy. Webster et al \cite{webster96} also provide a detailed analysis of the turbulence statistics of their experiment of a boundary-layer flow, but their adverse pressure gradient is created by a surface bump (quite similar to the one used in this paper) with concave and convex regions. In their experimental study of a boundary layer flow with an adverse pressure gradient, Dengel \& Fernholz~\cite{dengel90} compares different cases of pressure distribution, with and without reverse flow (transitory detachment in the classification of Simpson \cite{simpson81}). They show that the presence of recirculation leads to a significant change in the near-wall properties of the flow, even upstream of the reverse flow region. These different studies provided a useful description of the velocity fluctuation near the wall but the use of numerical simulations, which give all the information of the flow field, should allow to bring a new insight on wall turbulence with pressure gradient.

 One of the easiest way to introduce an adverse pressure gradient is to use suction at one wall of a plane channel flow or to directly prescribe the adverse pressure gradient. Na and Moin~\cite{na98a,na98b} performed a DNS of separated boundary layer flow on a flat plate using suction-blowing velocity distribution at the upper wall and second order finite difference scheme for spatial derivative. The inflow condition was taken from Spalart's temporal zero pressure gradient (ZPG) simulation. Chong et al~\cite{chong98} used the data from the DNS of Na \& Moin and from the Spalart's DNS of ZPG flow~\cite{spalart88} to analyze the topology of near wall coherent structures using invariant of the velocity gradient tensor. Spalart~\cite{spalart93} compared experimental and DNS results of a boundary layer flow with pressure gradient. The DNS of Spalart was performed using a spectral code with a fringe region in order to deal with periodic conditions in the non-homogeneous streamwise direction and a friction velocity at the edge of the boundary layer was prescribed to reproduce the pressure gradient of the experiment. More recently, Skote and Henningson~\cite{skote02} performed DNS of separated boundary layer flow  with two different adverse pressure gradients at Reynolds numbers $Re_\delta^*=400$ at the inlet (where $\delta^*$ is the momentum thickness).The advantage of those methods, with a prescription of the pressure gradient, is the possibility to use a numerical code adapted to plane boundary layers or plane channel flows. However the use of curved walls is more challenging. Some  recent simulations of such flows have been conducted with finite differences or finite volume codes  which are usually  less efficient than spectral codes to perform a highly resolved simulation. Neumann et al \cite{neumann04}) have investigated the effect of flow control on the  flow separation which requires a fine enough spatial discretization  to capture the smallest structures of the flow. It means that only DNS or highly resolved LES can be used. Wu et al \cite{wu98} have performed a LES of a boundary layer over a smooth bump and have compared the results to an experiment conducted earlier by Wester et al \cite{webster96}. However, the use of coarse resolution with an eddy viscosity model did not allow an accurate investigation of very small coherent structures close to the wall.

The numerical code used for the present study combines the advantages of the good accuracy of a spectral resolution and a fast integration procedure for simulations over a smooth profile. This code was originally developed to study the 2D and 3D instabilities of a boundary layers over a bump \cite{marquillie02,marquillie03}. It was here adapted to investigate the effect of a pressure gradient on the turbulent structures  at moderate Reynolds number: $Re=h U_{max} / \nu = 7900$ where $h$ is half the channel width and $U_{max}$ the maximum velocity at the inlet. At this Reynolds number the flow slightly separates on the bump but not at the opposite flat wall. This allows a comparison of the statistics of turbulence in the two configurations. The DNS of such a flow is very challenging because it requires both a very fine mesh in order to capture the smallest intense structures in the separation region, but also a long integration time to correctly access the slow oscillation of the separation line. In the present situation the flow does not exhibit a large separation bubble and the analysis is focused on the smallest structures instead of the mean flow characteristics.

In a first part, the numerical procedure will be described including the mapping of coordinates and the time integration. Then, the physical and numerical parameters of the DNS will be detailed. In a third part, the flow will be characterized in terms of average quantities and coherent structures. Some second order moments statistics of turbulence will be analyzed in the last part.

\section{The numerical procedure}

\subsection{Mapping and discretization}

Rather than writing the Navier-Stokes system in curvilinear coordinates we transform the partial differential operators using the following mapping (in order to simplify the equations in this section, the barred quantities will represent the physical coordinates):
\begin{equation} 
t = \bar{t}\:\: ,\:\: x = \bar{x}\:\: ,\:\: y 
= \frac{1}{L} \left(1-\gamma (\bar{x})\right) \bar{y} + \gamma (\bar{x})
\:\:,\:\: z = \bar{z} .
\label{eq:transform}
\end{equation}
with
\begin{equation} 
\:\:\:\: \gamma(\bar{x}) = \frac{L+\eta(\bar{x})}{\eta(\bar{x}) -L}
\end{equation}

The mapping used in \cite{marquillie02,marquillie03} (which was only a function of $x$) was suitable for a boundary layer flow but not for a channel flow. The new mapping used for the present simulation has the property of following the profile $\eta (x)$ at the lower wall with a flat surface at the upper wall. The transformed gradient and Laplacian operators are separated in two parts:
\begin{equation} 
\vec{\bar{\nabla}} = \vec{\nabla}_{\eta} + \vec{G}_{\eta} \;\;\;, \;\;\;
 \bar{\Delta} = \Delta_{\eta} + L_{\eta},
\label{gradop}
\end{equation}
the operator $ \vec{G}_{\eta}$ and  $L_{\eta}$ coming from the mapping are defined in the appendix A.
Consequently, the physical domain is transformed into a Cartesian one in the computational variables $(x,y,z)$. The three-dimensional Navier-Stokes system then writes
\begin{equation} 
\frac{\partial \vec{u}}{\partial t}  + ( \vec{u} . \vec{\nabla}_{\eta} ) \vec{u} +  ( \vec{u} . \vec{G}_{\eta} ) \vec{u}  =  - \vec{\nabla}_{\eta} p  - \vec{G}_{\eta} p  + \frac{1}{Re} \Delta_{\eta} \vec{u}  + \frac{1}{Re}  L_{\eta} \vec{u} 
\end{equation}
\begin{equation} 
\vec{\nabla}_{\eta} . \vec{u} = - \vec{G}_{\eta}.\vec{u}
\label{div}
\end{equation}
The Poisson equation for the pressure is obtained applying the divergence  operator to the momentum equations together with (\ref{div}) :
\begin{equation} 
\Delta_{\eta} p =  -  L_\eta  p  +  J(\,u,\,v,\,w) \:\: 
\label{press}
\end{equation}  
where J(\,u,\,v,\,w) is defined in the appendix A.\\

 For space discretization, fourth-order central finite differences are used for the second derivatives in the streamwise $x$-direction. All first derivatives of the flow quantities appearing explicitly in the time-advancing scheme and the first derivatives in $x$ are discretized using eighth-order finite differences. Chebyshev-collocation is used in the wall-normal $y$-direction. The transverse direction $z$ is assumed periodic and is discretized using a spectral Fourier expansion with $N_z$ modes, the nonlinear coupling terms being computed using the conventional de-aliasing technique with $M > 3 N_z /2 $.

For time-integration, implicit second-order backward Euler differencing is used; the Cartesian part of the Laplacian $ \Delta_{\eta}$ is taken implicitly whereas an explicit second-order Adams-Bashforth scheme is used for the operators $\vec{G}_{\eta}$ and $L_\eta$ as well as for the nonlinear convective terms. The three-dimensional system uncouples into $N_z$ two-dimensional subsystems and the resulting 2D-Poisson equations are solved efficiently using the matrix-diagonalization technique (The second-derivative operator in $y$ is diagonalized once for all and the system is hence transformed into a series of one-dimensional Helmholtz-like equations in $x$ which can be solved efficiently using the pentadiagonal structure of the matrices).

\subsection{Fractional-step method}

In order to ensure incompressibility, a fractional-step method  (cf. \cite{kim85}, \cite{karniadakis91}) has been adapted to our formulation of the Navier-Stokes system. The equations, once discretized in time, lead to the following system which has to be solved at each time step:
\begin{equation}
( \Delta_{\eta} - 3 \tau ) \vec{u}^{n+1} = Re  \vec{\bar{\nabla}} p^{n+1}    +  f^{n,n-1} \:\:\:  \mbox{with} \:\:\:  \tau = \frac{Re}{2 \Delta t },
\end{equation}
\begin{equation} 
\Delta_{\eta}  p^{n+1}  = [\, - L_{\eta} p + J(u,v,w) ]^{n,n-1},
\end{equation} 
\begin{equation}
\vec{\bar{\nabla}} . \vec{u}^{n+1}  =  0,
\end{equation}
with
\begin{equation}
f^{n,n-1} = - 4 \tau \vec{u}^n + \tau \vec{u}^{n-1} - [L_{\eta} \, \vec{u}\;   ]^{n,n-1} +  Re [( \vec{u} . \vec{\nabla}_{\eta} ) \vec{u} + ( \vec{u}  .  \vec{G}_{\eta} ) \vec{u} ]^{n,n-1}.
\end{equation}
Here the superscript $ n , n-1 $ means an explicit Adams-Bashforth differencing with $[.]^{n,n-1} = 2 \,[.]^n - [.]^{n-1}$. 
The outflow condition is 
\begin{equation}
\frac{\partial \vec{u}}{\partial t} + 
U_c  \frac{\partial \vec{u}}{\partial x} = 0,
\end{equation}
with $U_c$ the mean convective velocity at the outlet.

Although numerous papers have been devoted to different formulations of the fractional-step method (or projection method) we briefly summarize the formulation used for the present computations. Indeed, in our formulation we had to take into account the metric terms depending on $\eta $, that is the profile of the bump. Knowing the flow quantities at previous time steps $n, n-1$, an intermediate pressure is computed via 
\begin{equation} 
\Delta_{\eta}  p^* = [J(u,v,w)- L_{\eta} p ]^{n,n-1} 
\end{equation} 
with the Neumann-type boundary condition as following:
\begin{equation}
\vec{\nabla}_{\eta} p^*.\vec{n}= - \left[ \frac{3\vec{u}^{n+1}-4\vec{u}^n+\vec{u}^{n-1}}{2\Delta t} - \left[ (\vec{u}.\vec{\nabla}_{\eta})\vec{u} +( \vec{u}.\vec{G}_{\eta})\vec{u} \right. \right. 
\end{equation}
\begin{equation}
\left.  + \vec{G}_{\eta}p  + \frac{1}{Re}(\vec{\bar{\nabla}} \times(\vec{\bar{\nabla}} \times \vec{u} ) \left. ) \right]^{n,n-1} \right]. (\vec{n} + \vec{n}_{\eta}) - \left[\vec{\nabla}_{\eta} p \right]^{n,n-1} . \vec{n}_{\eta} \;. 
\label{neumnavier}
\end{equation}
Here we update at each time step the boundary condition for the pressure in order to minimize the effect of erroneous numerical boundary layers induced by splitting methods, as suggested by Karniadakis {\it et al.}  \cite{karniadakis91}. In a quite recent paper Hugues and Randriamampianina \cite{hugues98} showed that to update the pressure-gradient at the boundary provides indeed a time accuracy of the same order as that of the underlying  temporal scheme. The intermediate pressure once obtained, one recovers an intermediate velocity field solving
\begin{equation}
( \Delta_{\eta} - 3 \tau ) \vec{u}^* = Re \vec{\bar{\nabla}} p^* +  f^{n,n-1} \:\:  \mbox{with} \:\:  \vec{u}^*_{\Gamma} = \vec{u}^{n+1}_{\Gamma},
\end{equation}
$\Gamma$ being the boundary.

Our numerical experiments showed that for seek of stability (in particular for steeper bumps) it is much preferable to solve during the projection step the pressure correction $\phi=p^{n+1}-p^*$, with 
\begin{equation}
\vec{\bar{\nabla}} \phi = -\frac{3}{2\Delta t} (\vec{u}^{n+1}-\vec{u}^*) \:\: , \:\: \vec{\bar{\nabla}} . \vec{u}^{n+1}  = 0,
\label{eq:lap}
\end{equation}
implicitly in the physical gradient operator $\vec{\bar{\nabla}}$ containing both the Cartesian and the metric part. Applying the physical divergence operator to (\ref{eq:lap}), using the incompressibility condition, and imposing homogeneous Neumann boundary conditions for the pressure correction, one recovers the following iteration sequence
\begin{equation}
\Delta_{\eta} \phi^{k+1} = \frac{3}{2\Delta t}
\left(\vec{\nabla}_{\eta}.\vec{u}^* + \vec{G}_{\eta}.\vec{u}^*\right)
- L_{\eta} \phi^k \:\: \mbox{with} 
\end{equation}
\begin{equation}
\vec{\nabla}_{\eta} \phi^{k+1} . \vec{n}+
\vec{G}_{\eta} \phi^{k+1} . \vec{n}_{\eta} = 
\vec{\nabla}_{\eta} \phi^{k} . \vec{n}_{\eta}.
\end{equation}
In practice only few iterations are necessary in order to recover a pressure correction up to the second order in time which is the overall precision of the time-marching algorithm. Hence, the new pressure as well as a divergence-free velocity field is obtained writing:
\begin{equation}
\vec{u}^{n+1}=\vec{u}^* - \frac{2\Delta t}{3} (\vec{\nabla}_{\eta} \phi + \vec{G}_{\eta} \phi) \:\: \mbox{and} \:\: p^{n+1}=  p^* + \phi .
\end{equation}

\section{Description of the simulation}

\begin{figure}
\centerline{\includegraphics[width=0.9\columnwidth]{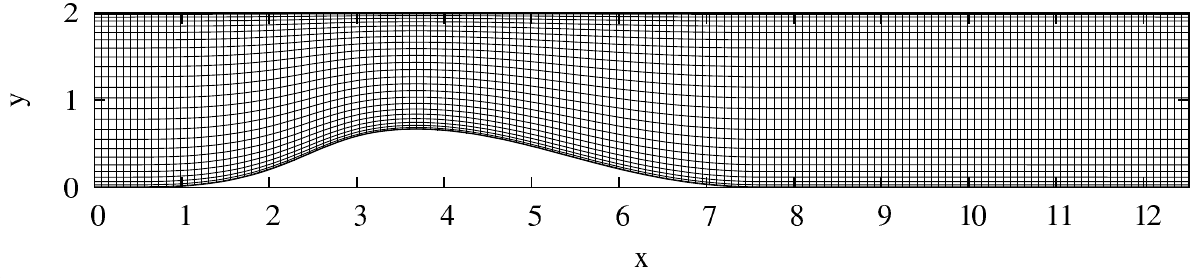}}
\caption{\label{fig:grid} Grid obtained after the transformation of coordinate defined by eq. (\ref{eq:transform}) (every 10 meshes are plotted in each direction).}
\end{figure}

We consider an incompressible fluid of viscosity $\nu$ flowing in a converging diverging 2D channel. The three-dimensional Navier-Stokes is made dimensionless using the half of the channel height $h$ as reference length, the maximum velocity at inlet $U_{max}$ being the reference velocity, the Reynolds number is:
\begin{equation} 
Re = \frac{U_{max} h}{\nu}.
\end{equation}
The lower limit of the domain is given by the profile of the bump $\eta (x)$ and the Navier-Stokes system has to be solved in the domain $0 \leq x \leq 4 \pi$, $\; \eta(x) \leq y < L$ where $L=2 h$ in our case . A local orthogonal coordinate system $(\vec{t},\vec{n},\vec{z})$ is introduced at the walls, $u_t$, $u_n$ and $u_z$ being respectively the tangential, normal and spanwise velocity components.

The bump was designed to be identical to the one used in the experiments conducted in the wind tunnel of the ``Laboratoire de M\'ecanique de Lille'' (see \cite{kostas05,bernard03}) at higher Reynolds number $Re_\tau \simeq 6500$  ($Re_\theta \simeq 20000$). The bump is characterized by a convex surface between $x=2.4$ and $x=5.4$ with two concave regions at the front and at the rear to achieve continuity with the flat wall. However, the simulation was not designed to reproduce the experiment as the Reynolds number of the simulation is one order of magnitude lower. The profile of the bump has been designed to have no separation at the Reynolds number of the experiment, but, as it will be shown, the flow slightly separates at the Reynolds number of the current DNS. The justification for the choice of the same profile is to reproduce a pressure gradient distribution comparable with the experiment.

\begin{figure}
\centerline{\includegraphics[width=0.7\columnwidth]{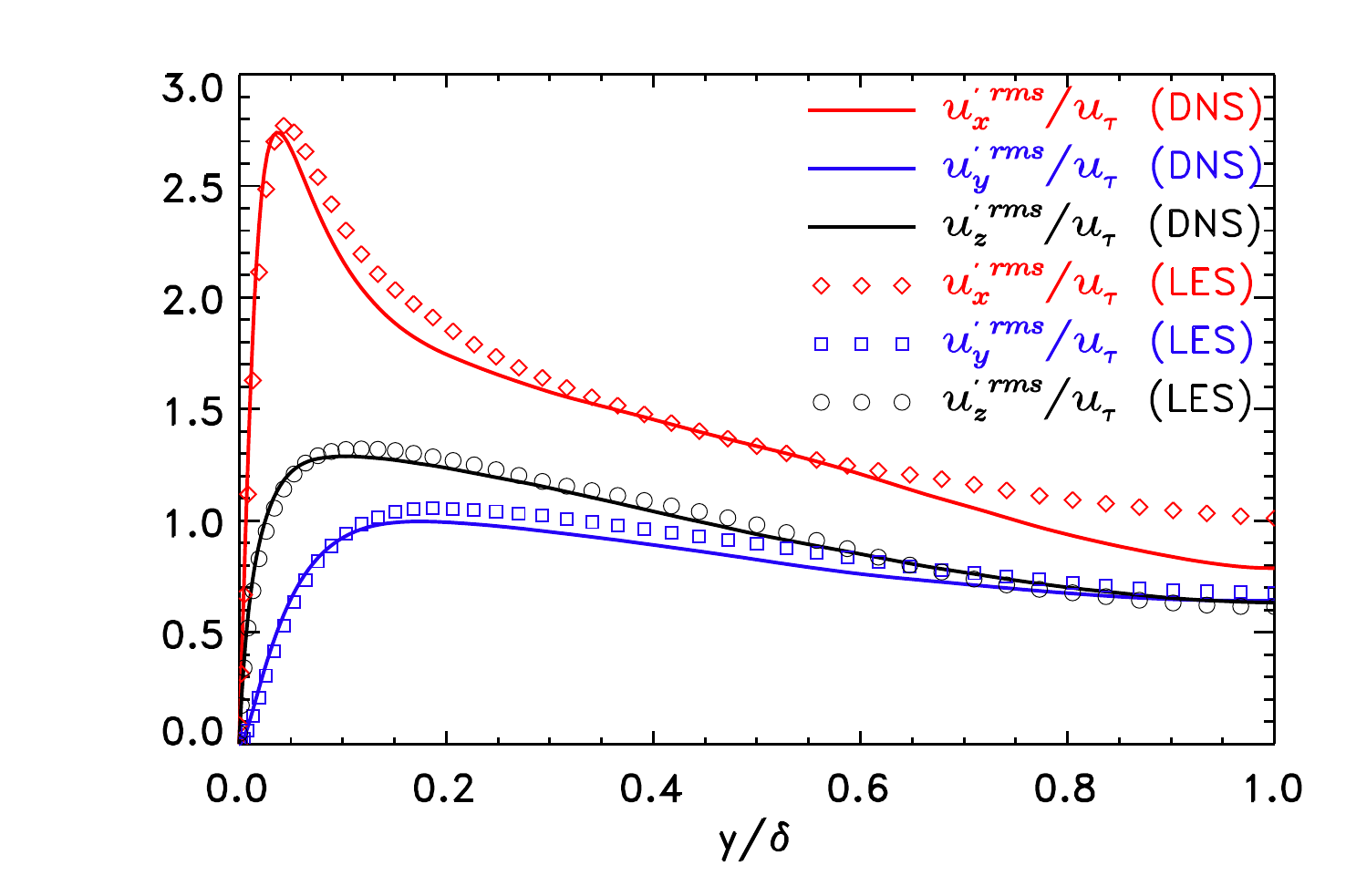}}
\caption{\label{fig:dnsles} Comparison of RMS of velocity fluctuations between the LES used as inlet condition with a DNS of Moser et al \cite{moser99}.}
\end{figure}

The whole domain of simulation is shown in Fig. \ref{fig:grid}. The definition of inlet boundary condition for DNS of turbulent flows which are not periodic in the streamwise direction is challenging and is still an ongoing research subject~\cite{moin98}. In order to bypass this difficulty the inlet conditions was taken from a previous highly resolved LES of plane channel flow at the same Reynolds number ($Re_\tau=395$). A standard Smagorinsky model was used as sub-grid scales model and the simulation was conducted in a domain of size $2\pi \times 2 \times \pi$ with a spatial discretization of $256 \times 97 \times 128$. Statistics from this LES are compared to the DNS of Moser et al \cite{moser99} at the same Reynolds number (see Fig. \ref{fig:dnsles}).

\begin{figure}
\centerline{\includegraphics[width=1.0\columnwidth]{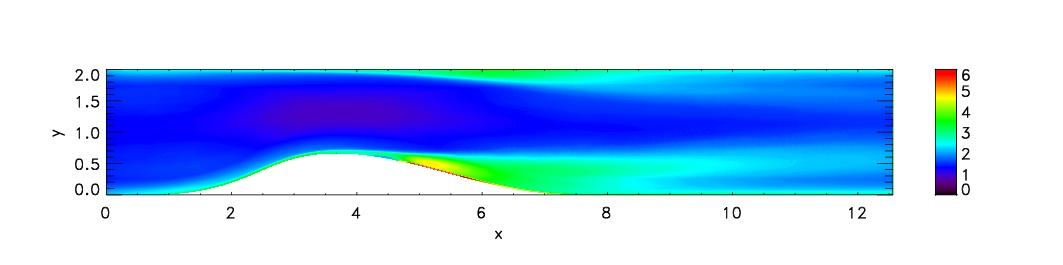}}
\caption{\label{fig:checkresol} Ratio of the maximum mesh size and Kolmogorov scale ($max(\Delta x,\Delta y,\Delta z)/\eta$) where $\eta=(\nu^3/\epsilon)^{1/4}$.}
\end{figure}

 The low performances of the Smagorinsky model for such a flow was compensated by a fine grid resolution. A good resolution of the LES was required in order to generate inlet conditions with correct coherent structures. For the DNS, due to the lower frequency response of the fourth order finite difference operator as compared to a spectral operator, the discretization in the streamwise direction has been doubled compare to the usual practice in channel flow simulation with fully spectral code. A second justification of the fine resolution in this direction is to correctly resolve the smallest structures created near the flow separation. A spatial resolution of $1536 \times 257 \times 384$ was finally used, which corresponds to a maximum mesh size of $3.9 \eta$ in the normal direction and $6.8 \eta$ in the two other directions where $\eta=(\nu^3/\epsilon)^{1/4}$ is the local isotropic Kolmogorov scale (see Fig. \ref{fig:grid}).  The ratio of the mesh size to the Kolmogorov scale is illustrated in Fig. \ref{fig:checkresol}. The maximum is localized in a thin layer at the lower wall in the separation region and a ratio from 3 to 5 is obtained in the regions of maximum Reynolds stress. The resolution in wall units based on $U_\tau$ at the inlet is $\Delta x^+\!\!=\!\Delta z^+\!\!=\!3, \; {\Delta y}_{max}^+\!\!=\!4.8$ and $\Delta x^+\!\!=\!\Delta z^+\!\!=\!7.6, \;  {\Delta y}_{max}^+\!\!=\!8.2$ based on the maximum value of friction velocity ($U_\tau=0.118$ at $x=2.93$). The time step is kept constant at a value of $4.0 \times 10^{-4}\;$s which corresponds to a CFL of approximately 0.15. The CFL is constrained by the high normal velocity induced by strong vortices close to the surface of the bump, downstream the separation region. The simulation was integrated over a physical time of 17.4 s which corresponds to approximately 1.6 flow-through time. Statistics of turbulence are converged after an integration time of approximately 9s. The simulation was initialized using 128 IBM SP4 processors at IDRIS (CNRS computing facilities) and followed with 128 IBM SP5 processors at CRIHAN (Center of Computing Ressources of Haute-Normandie, France).

\section{Results}

\subsection{Characterization of the flow}
\begin{figure}
\centerline{\includegraphics[width=0.7\columnwidth]{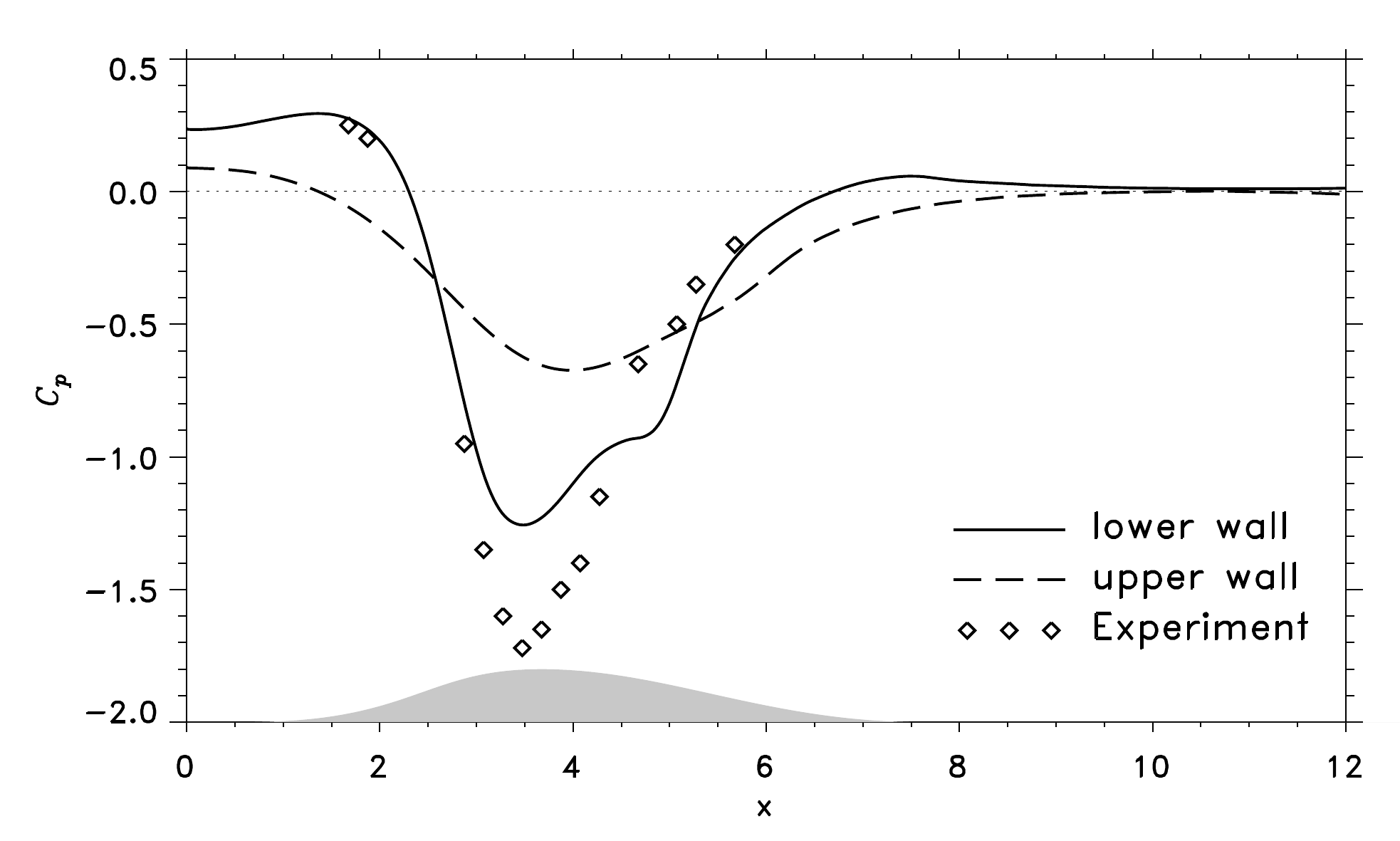}}
\caption{\label{fig:cp} Pressure coefficient at the two walls: $C_p = (P-Po) / \left( \frac{1}{2} \rho U_{max}^2\right)$ where $U_{max}=1.03$ is the maximum velocity at the inlet and $Po$ is a reference pressure near the outlet ($x=12,y=1$). The corresponding $Cp$ (using static wall pressure measurements) of the experiment of Bernard et al \cite{bernard03} for a boundary layer flow with no separation is given for comparison. The profile at the lower wall is plotted in grey as a reference.}
\end{figure}

 The pressure coefficient of the current DNS and the experiment of Bernard et al \cite{bernard03} are shown Fig. \ref{fig:cp}. The difference between the experiment and the DNS was expected due to the difference in Reynolds number and the channel flow inlet conditions of the DNS which are somewhat different from the experiment. The behavior in the converging part are similar but the minimum value at the summit of the bump is 30\% lower for the DNS as compare to the experiment.\\

\begin{figure}
\centerline{\includegraphics[width=\columnwidth]{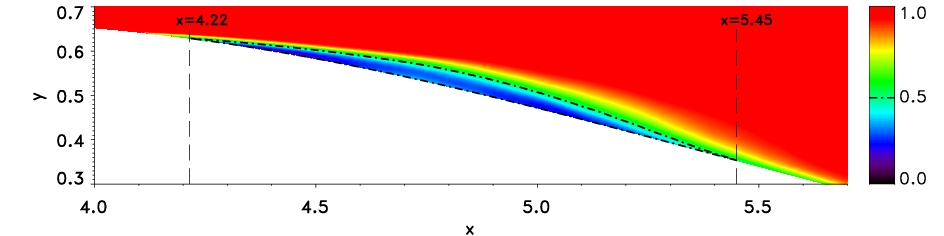}}
\caption{\label{fig:reverse} Probability of forward flows $\gamma_u$ in the separation region downstream the summit of the bump averaged in spanwise direction.}
\end{figure}

\begin{figure}
\centerline{\includegraphics[width=0.7\columnwidth]{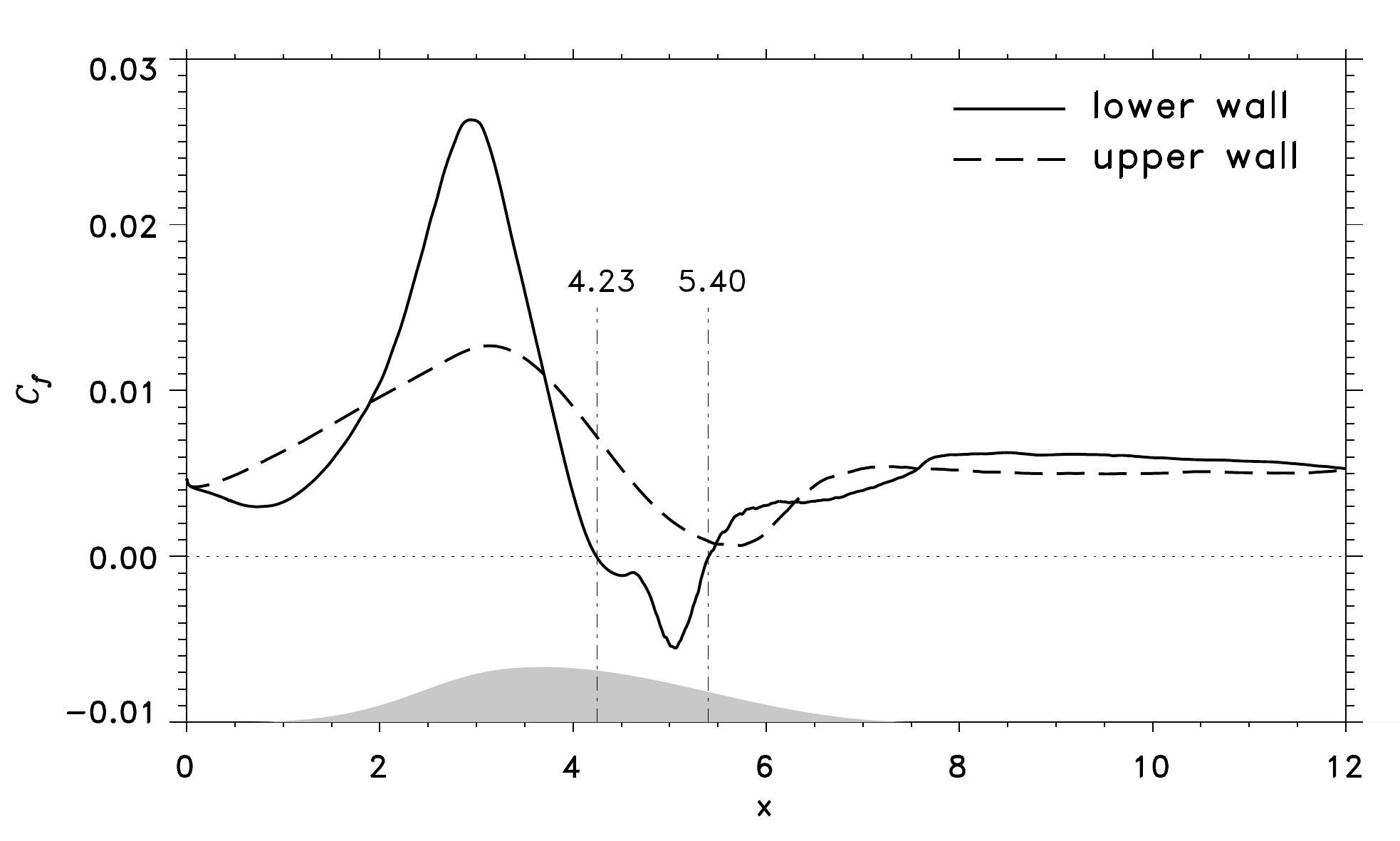}}
\caption{\label{fig:skin_friction} Skin friction coefficient $C_f = \tau_w / \left( \frac{1}{2} \rho U_{max}^2\right)$ (with $\tau_w=\mu \frac{d\langle u_t \rangle}{dy}\vert_{y=0}$) computed at the two walls with the mean tangential velocity $<u_x>$ averaged over $\Delta T=9s$ and the maximum velocity at the inlet ($U_{max}=1.03$). The separation region defined with negative shear stress is indicated with two vertical lines.}
\end{figure}

 In order to quantify the intermittency of the flow near the wall, the statistic of reverse flow occurrence is computed at each point and averaged in the spanwise direction and in time ($\gamma_u$ is the fraction of the time that the flow moves downstream). The result (Fig. \ref{fig:reverse}) exhibits a thin layer where the reversing flow conditions does exceed 50\% of time. A definition of separation and reattachement has been proposed by Simpson~\cite{simpson81} for separating turbulent boundary layer. He defined four different states: {\it incident detachment} which correspond to $\gamma_u = 0.99$, {\it intermittent transitory detachment} which occurs with $\gamma_u = 0.80$, {\it transitionary detachment} for $\gamma_u = 0.50$ and the {\it detachement} occurs when the time averaged wall shearing stress is zero. Using this definition, the region of transitory detachment does not exceed 10\% of the bump hight (see Fig. \ref{fig:reverse}) and the separation and reattachment points are located at $x=4.23$ and $x=5.4$ respectively (see the skin friction coefficient Fig. \ref{fig:skin_friction}). The streamwise size of transitory detachment region at the wall is almost identical to the size of the separation.  The strong decrease of the skin friction $C_f = \nu \frac{d<u_t>}{dy}\vert_{y=0} / \frac{1}{2}U_{max}$ near $x=4.5$ coincides with the increase of normal velocity flutuation. This seems to indicate that the intense vortices evolving close to the bump strongly affect the skin friction. As shown in Fig. \ref{fig:skin_friction}, the skin friction reaches a very small value at the flat upper wall which indicates that the flow is very closed to separation.

\subsection{Vortices in separation region}

\begin{figure}
\centerline{\includegraphics[width=\columnwidth]{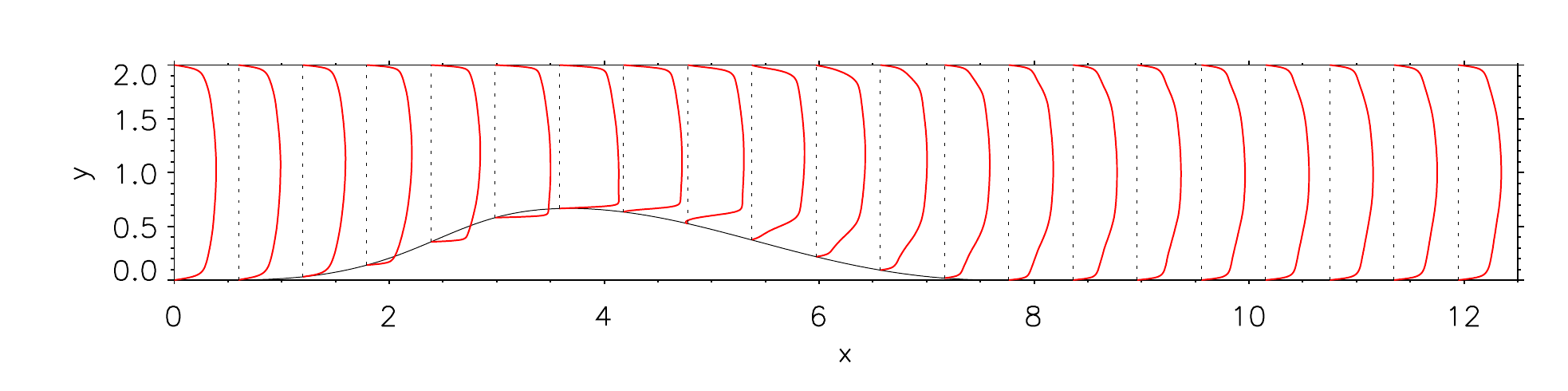}}
\caption{\label{fig:umean} Profiles of mean streamwise velocity $\langle u_x \rangle$.}
\end{figure}

\begin{figure}
\caption{\label{fig:umean2} Profiles of mean tangential streamwise velocity $\langle u_t \rangle$ at the lower curved wall (left) and upper flat wall (right).}
\end{figure}

Several profiles of the mean streamwise velocity are plotted in Fig. \ref{fig:umean}.
The mean velocity profile at the outlet does not recover the same shape as at the inlet. As it will be discussed later, the turbulence intensity does not recover its inlet values as well. When computing the mean of the tangential velocity (see Fig. \ref{fig:umean2}), the profiles exhibit a very thin separation bubble at the lower wall in the downstream part of the bump ($\delta n \simeq 0.02$ at $x=5$). A inflexion point in the profile is already perceptible near the summit at $x=4$. The size of the near wall layer decreases significantly over the bump. However, as the inlet conditions are generated from a channel flow, it is not possible to accurately define a boundary layer as it is usually defined for a pure boundary layer flow. The same profile at the upper wall confirms that there is no separation. However the mean velocity profile at $x=5.5$ also exhibit a clear inflexion point, which is generally associated to convective or absolute instabilities in transitional flow.\\

\begin{figure}
\centerline{\includegraphics[width=0.95\columnwidth]{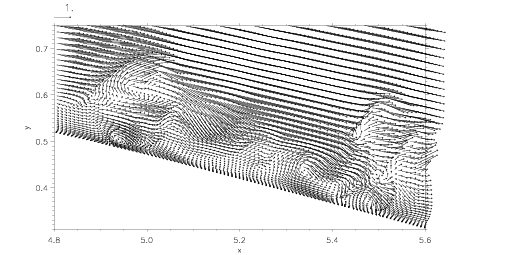}}
\caption{\label{fig:velocity} Example of intense vortices near the wall in a sample of velocity vectors within the separation region (every other vectors are plotted in each direction).}
\end{figure}

\begin{figure}
\centerline{\includegraphics[width=\columnwidth]{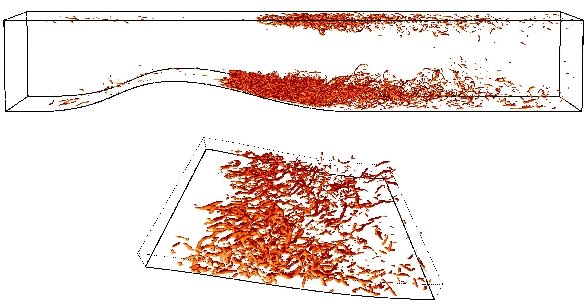}}
\caption{\label{fig:structure} \label{mov:structures1} Isovalues of the Q invariant of the velocity gradient tensor ($Q=\Omega^2-S^2$) in the full simulation domain and in the diverging part at the lower wall ($4.5<x<7.5$).}
\end{figure}

In order to study the global motion of the turbulent structures, isovalues of the second invariant $Q=\Omega^2-S^2$ of the velocity gradient tensor are presented in Fig. \ref{fig:structure} and in the associated movies 1 and 2. The vortices generated near the separation are much more intense than the coherent structures generated in a plane channel flow. Consequently, the isovalue used to detect the vortices downstream of the bump is not adapted to detect structures at the inlet of the simulation. The generation of intense coherent structures is nearly steady in time and in space, but the localization is different at the two walls and is slightly downstream of the position of the first inflexion point of the mean velocity profile ($x \approx 4.8$ at the lower wall and $x \approx 5.8$ at the upper wall). However, the typical size of the vortices seems comparable at both walls and vary very slowly in time. This behavior may be due to the detection criteria which only capture well formed and intense vortices.


A 2D slice of the time evolution of $Q$ in the separation region enable us to have a closer look of the generation of vortices. The separation region can be divided in two parts. Almost no vortices are detected in the upstream part ($ 4.2 < x < 4.8$) but strong ones are generated close to the wall in the downstream part of the separation region. These vortices interact with those convected in the outer boundary layer. This leads to non trivial motions of these small near wall vortices. Only a small part of them are convected by the mean reversal flow very close the wall but most of them are either destroyed by the shear or by interaction with larger vortices. As shown by the instantaneous velocity field  (Fig. \ref{fig:velocity}), some vortices lift up and occasionally generate an ejection of fluid in the outer region of the flow.   \\

\subsection{Near wall structures}

\begin{figure}
\centerline{\includegraphics[width=\columnwidth]{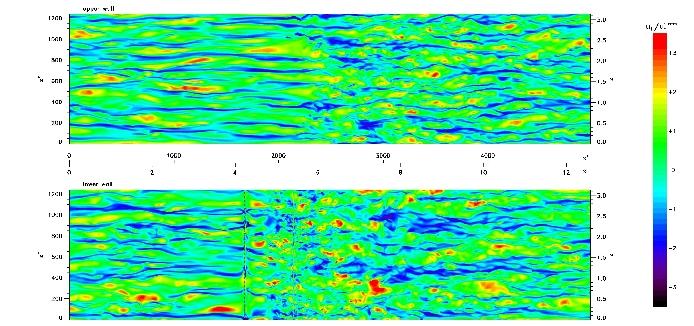}}
\caption{\label{fig:streaks} \label{mov:streaks} Iso-contours of streamwise normalized velocity fluctuation interpolated in a plane at $n^+=n U_\tau(x)/\nu = 10$ from the lower wall and upper wall. The streamwise and spanwise coordinate have been non-dimensionalized by $U_\tau$ at the inlet ($U_\tau=0.0501$).}
\end{figure}

\begin{figure}
\centerline{\includegraphics[width=0.8\columnwidth]{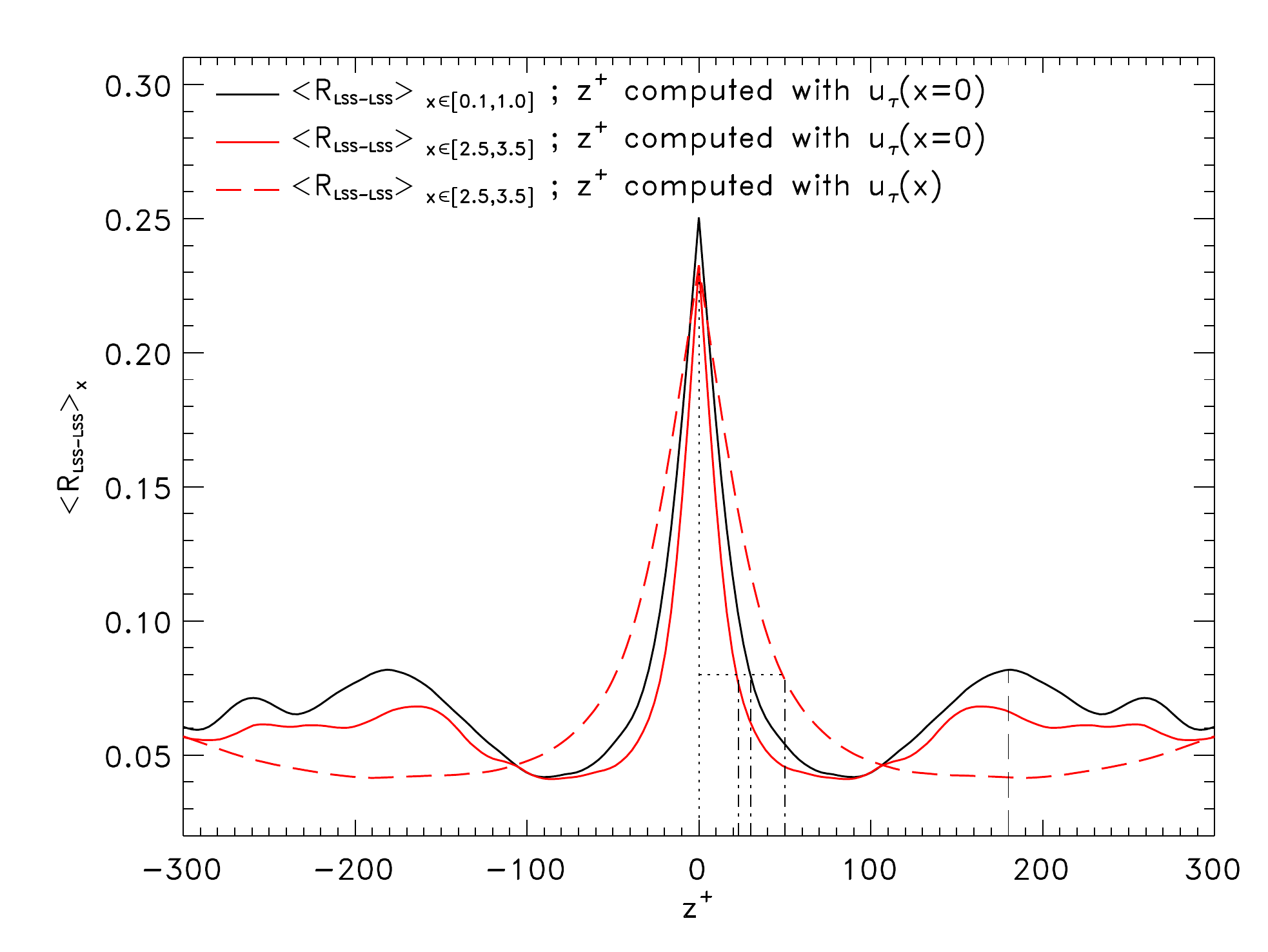}}
\caption{\label{fig:correl_streaks} Spanwise correlation of binary images of low speed streaks ($R_{LSS_LSS}$) extracted  with a detection procedure defined by Lin (\cite{lin06} and averaged in the inlet part (black curve) and converging part of the domain (red curves) at a distance $n^+=10$ from the lower wall.  The inlet value of the friction velocity was used to convert the spanwise coordinates into wall units for the two first curves (full lines). For these two first curves, a constant value of $U_\tau$ was used in order to compare the physical size of the streaks. For the third curve, in the converging part, the local value of $U_\tau$ was used before taking the average in order to compare the statistics in local wall units. The dot-dash lines indicates an estimation of the size of the  low speed streaks and the dotted line (located at the second  peak of correlation) indicates an estimation of the averaged distance between them.}
\end{figure}

The effect of the pressure gradient on the near wall streaks is illustrated in Fig. \ref{fig:streaks}. The fluctuating streamwise velocity normalized by its local standard deviation is interpolated at a constant distance of $\delta n^+=10$ from the two walls. Streaks are usually defined as the regions with $\vert u_t^{\prime}/u_t^{\prime \; rms} \vert > 1$ where $u_t^{\prime}= u_t - \langle u_t \rangle$ and $u_t^{\prime \; rms}= \sqrt{\langle u_t^{\prime 2} \rangle}$ are respectively the fluctuations and the standard deviation of the tangential streamwise velocity. In  figure  \ref{fig:streaks}, the spanwise and streamwise coordinates in wall units were computed with $U_{\tau}$ at the inlet but the local standard deviation was used to normalized the fluctuation. The size and the spacing between low speed streaks are difficult to access because of the small number of uncorrelated sample in the near wall region. The two-point correlation of normalized streamwise velocity fluctuation will be two noisy to extract the statistics of streaks. Hence, a detection procedure developed by Lin ~\cite{lin06} to extract the low speed streaks has been applied. The structures are detected at the lower wall using a thresholding of a detection function followed by a cleaning procedure based on image processing. Then, a correlation was computed in the spanwise direction from the binary images obtained from this detection procedure. The results of this correlation is shown Fig. \ref{fig:correl_streaks}. Because this correlation is more converged than the correlation of fluctuating velocity, the typical spanwise size and spanwise distance between low speed streak can be extracted from the first and second peak of correlation. The statistics were computed near the inlet ($0.1<x<1.0$) and in the converging part ($2.5<x<3.5$). This figure demonstrated that the low speed streaks are stretch (in physical unit) in the converging part. The spanwise size decreases from approximately 30 to 25 wall units based on $U_{\tau}$ at the inlet. Skote et al~\cite{skote02} observed similar behavior of low speed streaks in a DNS of boundary layer flow with adverse pressure gradient. However, when computed with the local wall units, the size increases up to approximately 50 wall units. This indicates that the low speed streaks are not able to adapt to the strong variation of friction velocity due to the pressure gradient. The size of the simulation domain is not large enough to compute accurate statistics on streamwise size of streaks. In zero pressure gradient flows they can reach several thousands of wall units which is comparable with the size of the bump. Near the summit of the bump  ($x^+ \simeq 3.8$), low and high speed streaks are still present but they are shorten by the separation ($x^+ \simeq 4.23$). Na \& Moin~\cite{na98a} pointed out the destruction of streaks in the separation region of their DNS of highly separated boundary layer flow and we observe similar behavior in our DNS with a thin separation region. On the upper wall, the situation is similar but less marked in the converging part. In the divergent part, after the region of negative friction velocity at lower wall (and low friction velocity at upper wall), streaks appear after $x^+=3000$ along both wall, but with a perceivable difference in structure between the upper and lower wall. This corresponds to approximately 1000 to 1500 wall units after the reattachment location on the lower wall and 500 wall units after the very low $U_\tau$ region on the upper wall. Upstream $x^+=3000$, short low speed streaks are still visible, but the turbulence activity appears more isotropic, more intense and at a smaller scales.

\subsection{Turbulence statistics}

\begin{figure}
\centerline{\includegraphics[width=\columnwidth]{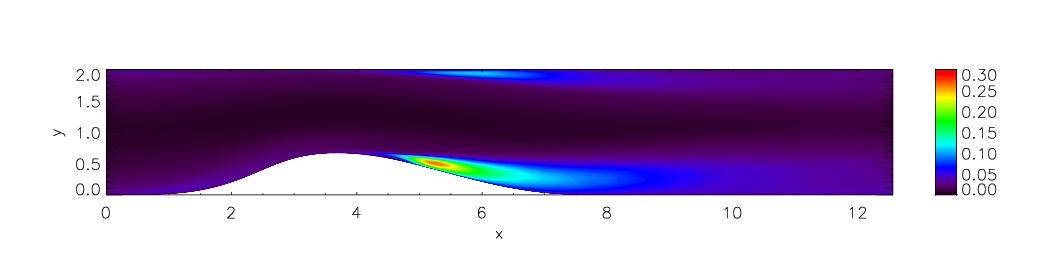}}
\caption{\label{fig:nrj} Turbulent kinetic energy normalized by the bulk velocity ($k/U_b^2$) in the whole simulation domain averaged over $\Delta T=9s$.}
\end{figure}

Contours of turbulent kinetic energy $k$ are displayed in figure \ref{fig:nrj} for the whole XY domain. The figure exhibits a maximum slightly downstream the summit of the bump which corresponds to the region of detection of intense vortices. The turbulent kinetic energy is much lower at the upper wall. The region of high turbulent kinetic energy extends up to $x \simeq 10$ on both side. In the converging part of the bump ($2.5<x<3.5$), $k$ is significantly reduced as compared to the plane channel and the same behavior is observable at the same location on the upper wall. The reduction of the boundary layer thickness concentrate $k$ very close to the wall, as a consequence the dissipation of the leading streamwise component of $k$ dominates the production.

\begin{figure}
\centerline{\includegraphics[width=0.8\columnwidth]{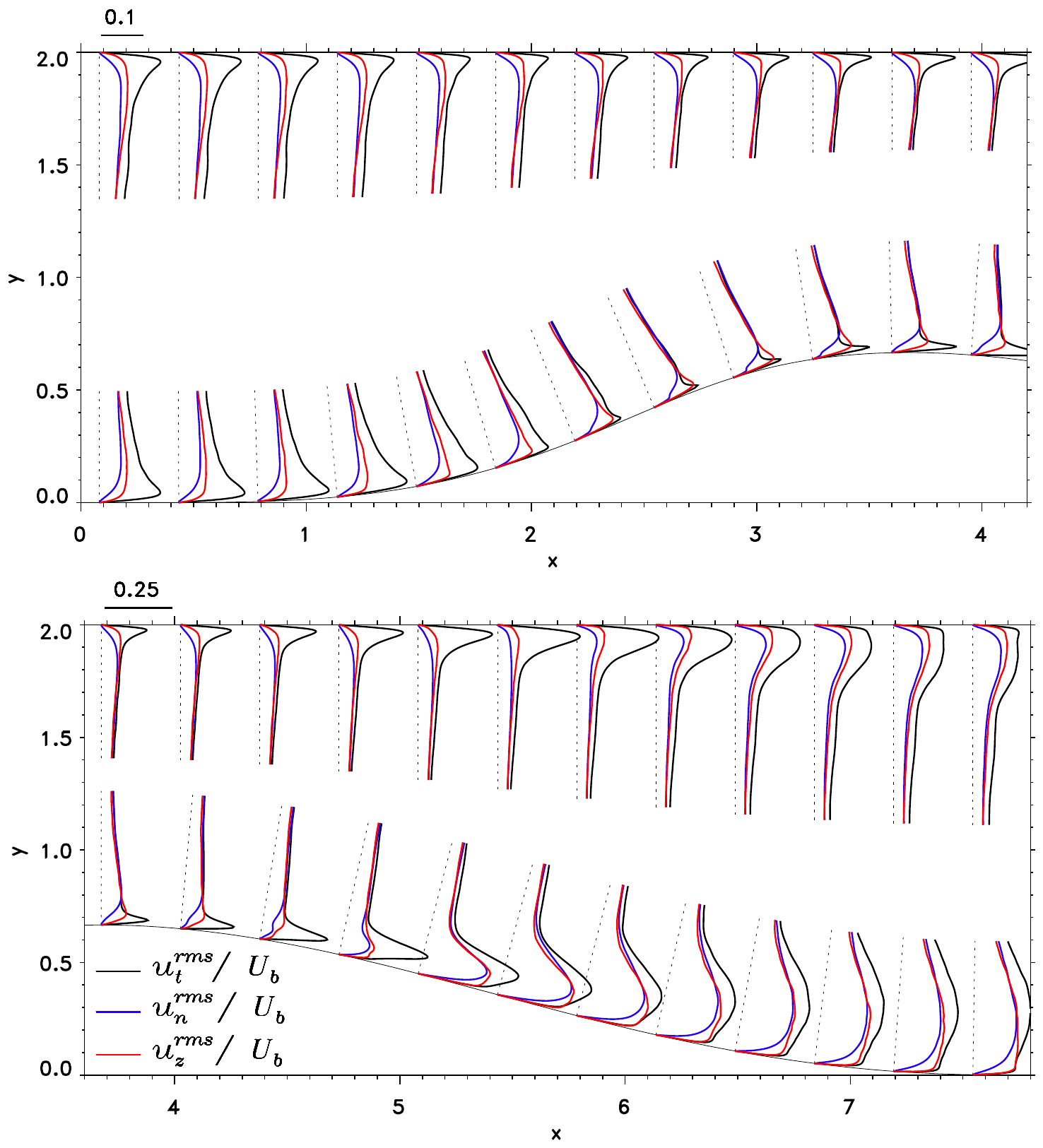}}
\caption{\label{fig:rms_profile} Profiles of the rms velocity fluctuation in the local coordinate system ($\vec{u} = u_t \, \vec{t} + u_n  \, \vec{n} + u_z  \, \vec{z} \;$ where $\vec{t}$ and $\vec{n}$ are the tangential and normal vectors respectively). The velocity profile have been non-dimensionalized by the bulk velocity $U_b=0.9$. The maximum value of $u_t^{\prime \; rms}/U_b$ at the station $x \simeq 5.1$ on the bump is 0.365.}
\end{figure}

Figure \ref{fig:rms_profile} shows the standard deviations of $u_t$, $u_n$ and $u_z$ over the bump and at the opposite wall. In the converging part, near wall peak of $u_t^{\prime \; rms}$ is comparable in magnitude at the lower and the upper walls. This is not true for the spanwise and normal components, which reach a  maximum value 50\% lower at the flat upper wall. In the converging part ($2<x<3$), the streamwise turbulence intensity diminishes significantly, both near the wall, where the peak is much smaller, and away from the wall, where a clear tendancy to isotropy appears and stays very far downstream ($x \simeq 6$ at the lower wall and $x \simeq 5.5$ at the upper wall).
At the summit of the bump ($x \simeq 3.5$), the profile of $u_t^{\prime \; rms}$  exhibits a very strong peak close to the wall. The two other components are much lower and their maxima occur at a distance from the wall comparable to the one observed uphill near $x \simeq 1$. The situation is similar at the upper wall, even if the peak of $u_t^{\prime \; rms}$ is less stretched near the wall. As reported by several authors for boundary layer flows with adverse pressure gradient \cite{webster96,aubertine05}, the profiles exhibit a second peak. In the present simulation, a new peak first appears close to the wall on the normal component at the separation location ($x \simeq 4.23$) because of the low amplitude of  $u_n^{\prime \; rms}$ near the wall. This new created peak grows rapidly in the recirculation region ($ 4.5<x< 5.5$) and extends in the normal direction. The behavior is similar for the spanwise velocity fluctuation but the new created peak appears slightly downstream because of the higher intensity of this component near the wall before the separation. The effect of the separation on the streamwise velocity fluctuation is only to increase the original peak near the wall. In the diverging part of the bump, the main peak of each component moves away from the wall indicating that the internal layer grows toward the outer layer. The new peak of $u_t^{\prime \; rms}$ and $u_z^{\prime \; rms}$ appearing near $x>6$ is the evidence of a new boundary layer starting to build up after the reattachment. The evolution of the stress components are similar on the upper wall, but the appearance of the second peak near the wall are delayed following the minimum of the skin friction. Also, the intensity of the phenomena is less pronounced as compare to the lower wall.

\begin{figure}
\centerline{\includegraphics[width=0.8\columnwidth]{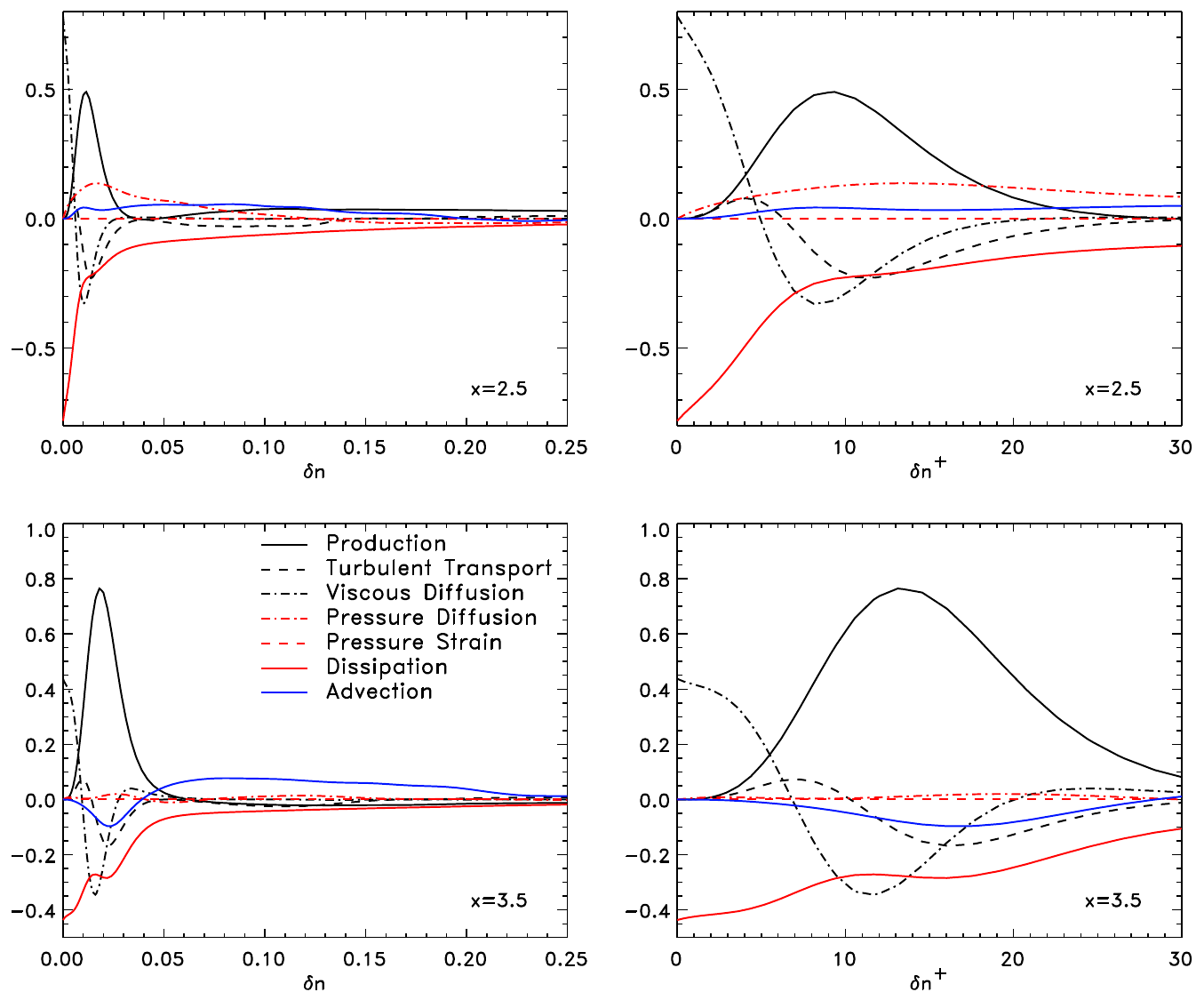}}
\caption{\label{fig:budget3} Budget of the turbulent kinetic energy (normalized by $U_\tau^4/\nu$ at $x=0$) at taken at 2 locations normally to the bottom wall in the converging part. The coordinates are in physical units (right) and in wall units (left) where the wall units are computed with the local friction velocity ($\delta n^+=\delta n \; U_\tau(x)/\nu$). The legend is identical for all the graphics.}
\end{figure}

 In an attempt to better characterize the turbulence evolution along the channel, the full budget of the Reynolds Stress equations was computed. The derivation and the definitions are given in  appendix B. The full budget will not be detailed here. As a representative example, the budget of the turbulent kinetic energy equation $k$ is presented. The budget is first described in the converging part at the lower wall (fig. \ref{fig:budget3}). As it has been previously mentioned, the peak of the streamwise component of the Reynolds Stress is significantly reduced in this region ($2.5<x<3.5$). This decrease is consistent with budget of $k$ which exhibit a high dissipation at the wall and an excess of dissipation as compared to the production in the near wall region ($0<n^+<30$).

\begin{figure}
\centerline{\includegraphics[width=0.8\columnwidth]{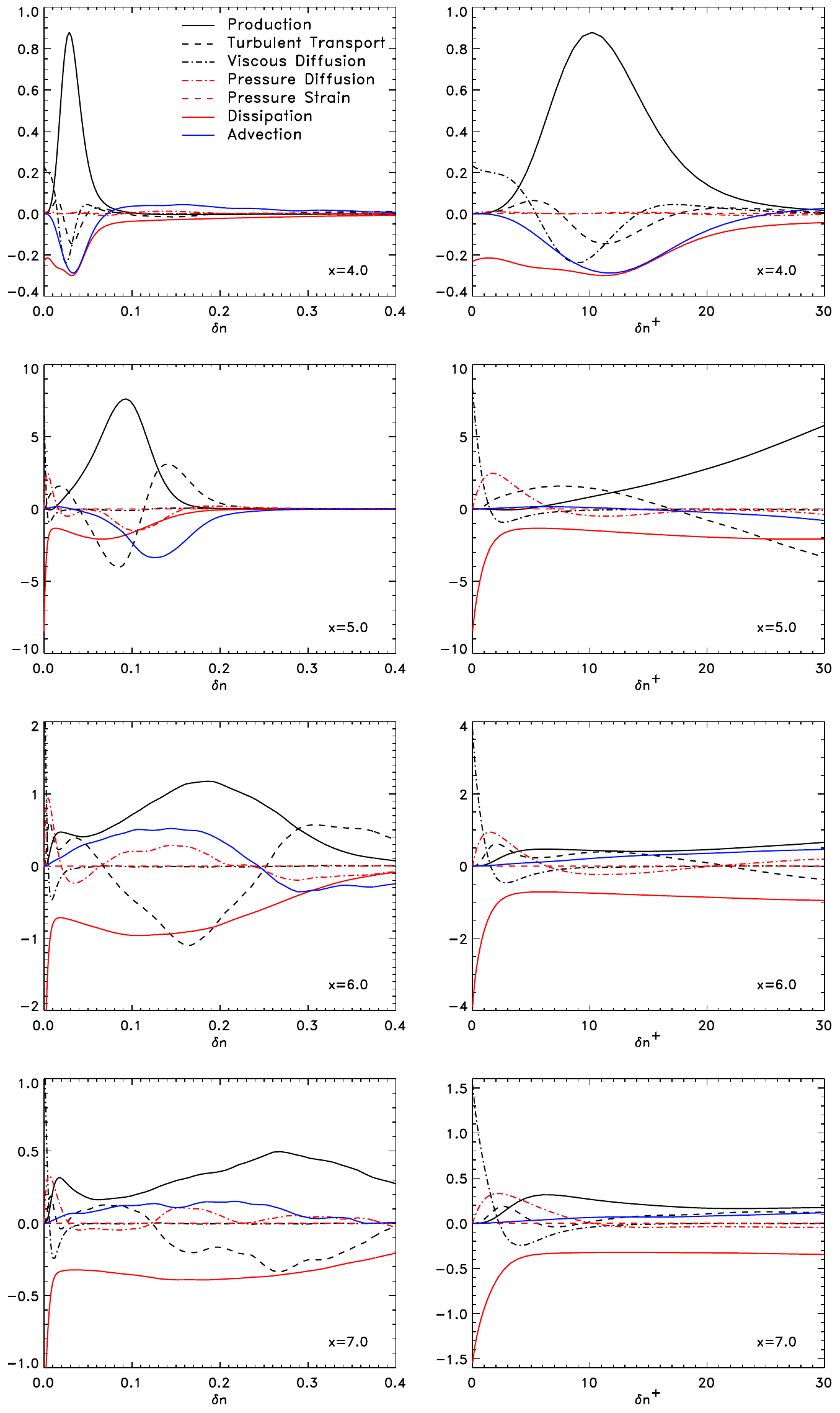}}
\caption{\label{fig:budget1} Same caption than Fig. \ref{fig:budget3} for 4 locations normally to the bottom wall.}
\end{figure}

 The budget of $k$ is also presented  in the diverging part at 4 locations for the lower wall (see fig. \ref{fig:budget1}). Near the summit of the bump ($x=4$), the peak of kinetic energy production is located at a distance $\delta n=0.03$ ($\delta n^+=10$), which corresponds to the starting point of the high values of $k$ (see Fig. \ref{fig:nrj}). At this location dissipation is lower than production and advection is significant.

The maximum peak value of the kinetic energy production, along the bump, is obtained at $x=5.0$. This peak corresponds to the minimum of the skin friction coefficient $C_f$. The peak is located at $\delta n=0.09$ which shows that the production moves away from the wall. At this location, production is approximatively 4 times higher than dissipation, this excess of produced energy  is transported away as shown by the advection (similar at the advection at $x=4$) and also by the turbulent transport. The turbulent transport is separated in three parts: a lost of energy where the peak of production is located, and a gain of energy towards the wall and towards the outer layer.

\begin{figure}
\centerline{\includegraphics[width=0.8\columnwidth]{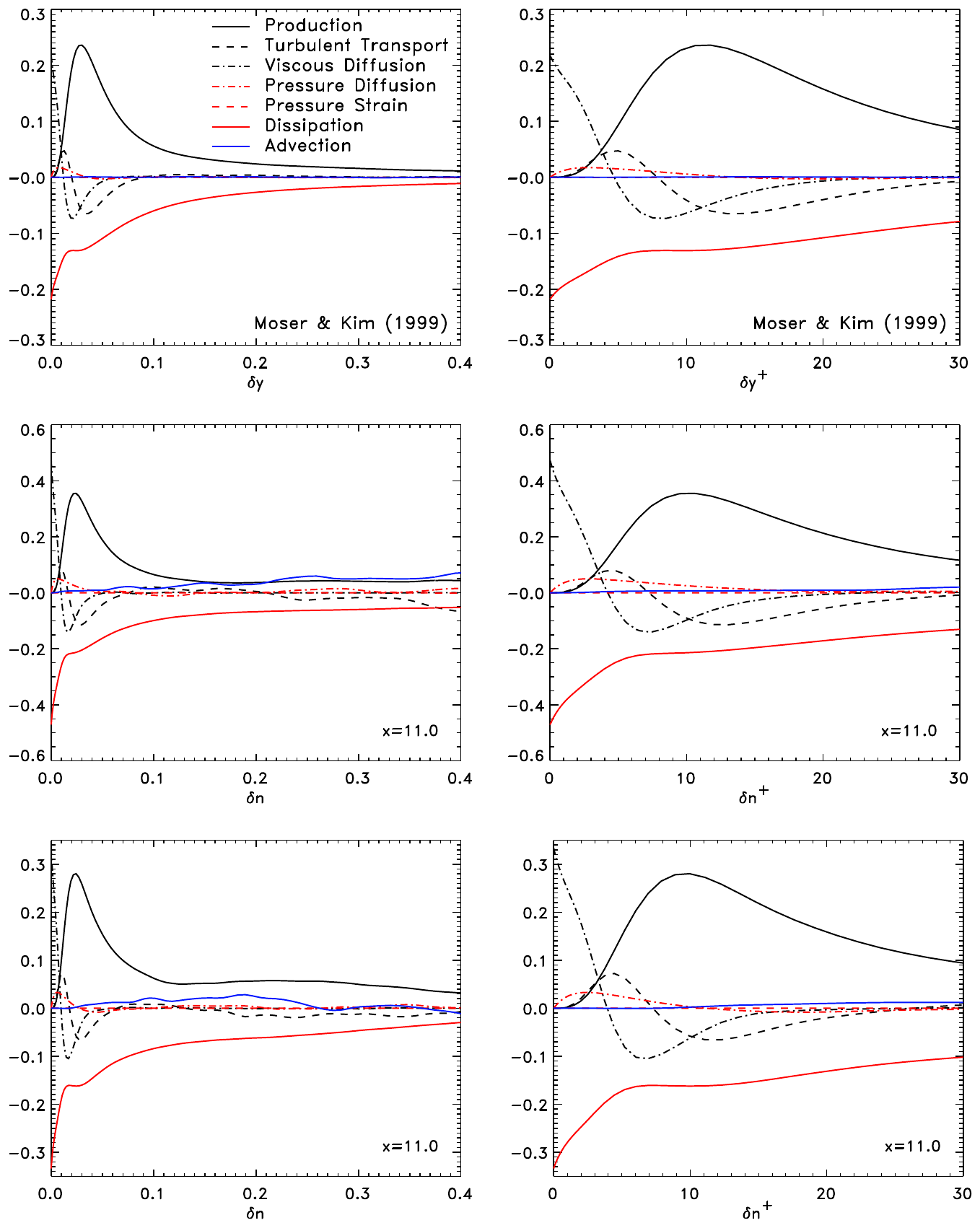}}
\caption{\label{fig:channel} Budget of the turbulent kinetic energy at $x=11$ for the lower wall (lower plots) and the upper wall (middle plots) and the same budget for a DNS of flat channel flow \cite{moser99} at $Re_\tau=392.24$ (upper plots) (from the AGARD database: {\sl ftp://torroja.dmt.upm.es/AGARD})}
\end{figure}

After the reattachment point ($x=6$), the production exhibits two peaks. The second peak (farther from the wall) is the peak observed in $x=4$ and $x=5$ which continues to move away from the wall at  $0.1<\delta n<0.3$ where most of the intense coherent structures are observed, but this peak in production has started to decrease significantly as compare to the high value observed at $x=5$. The first peak, lower in intensity, is located at the same physical distance to the wall ($\delta n=0.03$) which corresponds to the position of the peak of production in a zero pressure gradient channel flow. Turbulent transport has a similar behavior as at $x=5$, except for the intensity. Advection is also moved away from the wall but with a positive peak value near the wall. In their turbulent boundary layer near separation  experiment, Skare \& Krogstad~\cite{skare94,krogstad95} have observed these two peaks of production, one in the near-wall region and one in the outer layer. They also reported the same typical shape of turbulent transport with 3 peaks (positive-negative-positive), which is different from the shape with two peaks (positive-negative) observed in zero pressure gradient channel flow. Farther downstream, at $x=7$, the second peak of production and the dissipation have extended away from the wall and both decreased in magnitude following the tendency at $x=6$. In the diverging part, the dissipation, viscous diffusion and pressure diffusion are very large close to the wall as compare to the flat channel flow (see Fig \ref{fig:channel}). The same behavior was observed by Na \& Moin in a DNS of separated turbulent boundary layer~\cite{na98a}.

\begin{figure}
\centerline{\includegraphics[width=0.8\columnwidth]{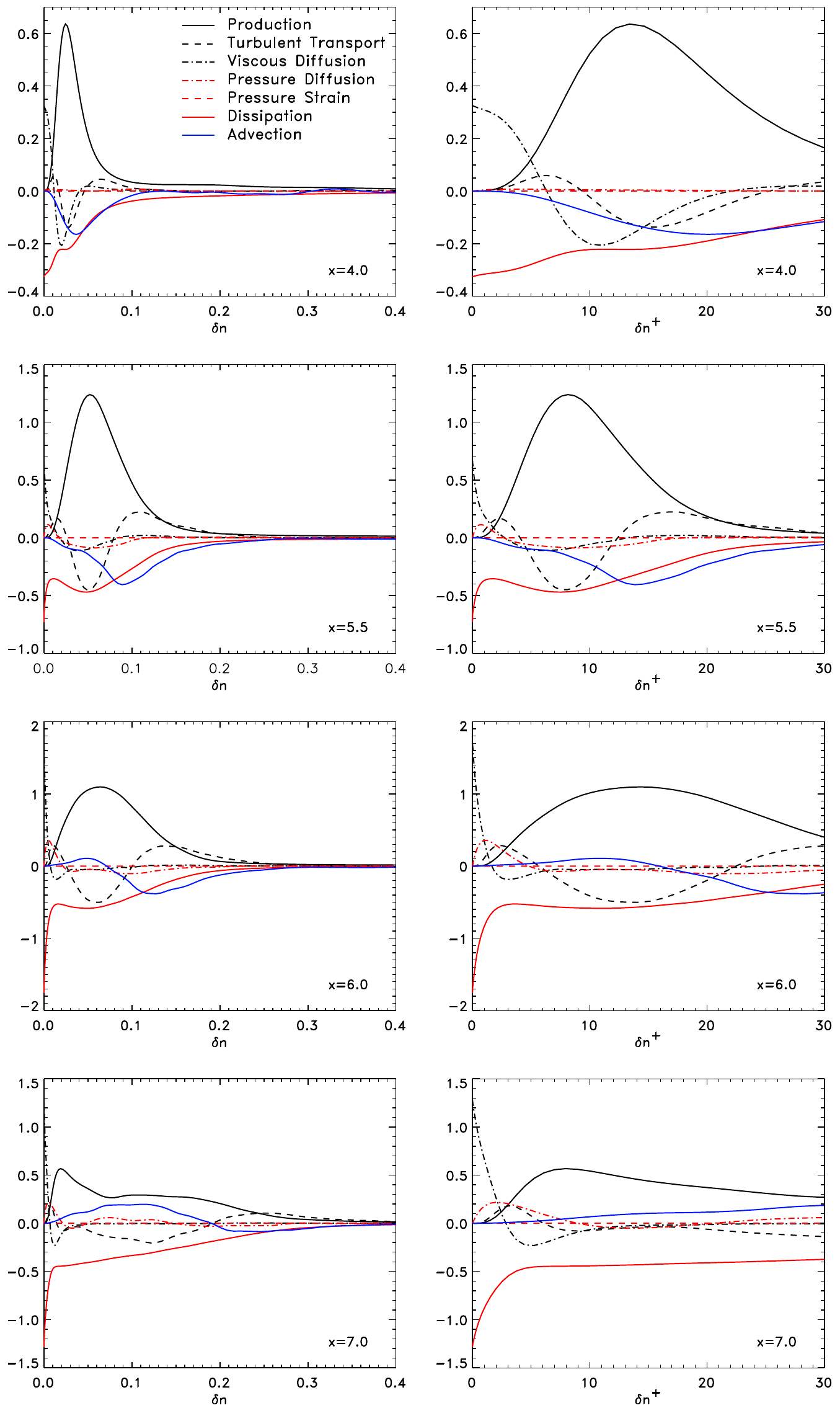}}
\caption{\label{fig:budget2} Same caption than Fig. \ref{fig:budget1} for upper wall.}
\end{figure}

The same budget is presented in Fig. \ref{fig:budget2} for the upper wall. The streamwise evolution of the different budgets terms for the upper wall is relatively similar to the evolution at the lower wall. The maximum peak of production (located at $x=5$ at the lower wall) is located slightly downstream at $x=5.5$ which corresponds to the position of the minimum of the skin friction. It should be noted that the increase of production at the upper wall is significantly lower than at the lower wall. The two peaks of production can also be observed at the upper wall at $x=7$. The first peak is comparable in magnitude with the first peak at the lower wall, but the difference comes from the second peak which is significantly smaller at the upper wall when it is still important at the lower wall. This explains the wider region of intense turbulent kinetic energy at the lower wall.

The budget of $k$ is presented (Fig \ref{fig:channel}) at $x=11$, near the end of the simulation domain but far enough from the outlet plane not to be influenced by outlet boundary conditions. The distribution of each terms of the balance is similar at the two walls with 30\% lower values of the dissipation and viscous diffusion at the upper wall. The budget of $k$ provided by Moser et al \cite{moser99} for a channel flow at the equivalent Reynolds number ($Re_\tau=392$) is also given in Fig. \ref{fig:channel} for comparison. Each component of the balance of $k$ at the end of our simulation domain has the same general behavior as for channel flow. However, the dissipation and viscous diffusion at the upper wall are still  30\% larger than the same statistics for the channel flow. The same behavior is observable for the peak of the turbulent transport and the peak of production with an over prediction of respectively 30\% and 10\% compare to the channel flow. The modification of the turbulence created by the adverse pressure gradient is strong enough to modify the balance of $k$ up to 10 bump height after the reattachement.

\clearpage

\section{Conclusion}

A DNS of channel flow in a smooth converging diverging channel has been presented. The fact that this single configuration is able to account for an adverse pressure gradient flow with and without curvature at the same time makes it an ideal test case for engineering.  The channel flow inlet condition are well documented and easy to model for simulations with a turbulent model like Reynolds Averaged Navier Stokes or LES models. The numerical code used for the present simulation is original as it uses a transformation of coordinates. This numerical code allows us to perform simulations with smooth profiles at reasonable cost, keeping the benefit of the efficiency and the accuracy of the spectral solver in spanwise and normal direction.

 The  flow slightly separates on the curved lower wall but not at the opposite flat wall which allows a comparison of the statistics of the turbulence in both situations.  The present study point out the occurrence of strong coherent structures generated near the separation region on the lower wall but also at the opposite wall where no noticeable separation occurs. At the lower wall, some small vortices are created in the downstream part of the separation region and interact with larger ones convected in the outer boundary layer. However, the beginning of the separation region presents no discernible vortices. This is confirmed by the low level of turbulent kinetic energy in this region. 

The low speed streaks are stretched by a favorable pressure gradient and disappear in the detachment region. The breakdown of streaks may be the consequence of strong vortices with downward motion. The streaks are regenerated approximately one thousand of wall units downstream the reattachment point. 

The turbulent kinetic energy is slightly lower at the upper wall with a different distribution between the three velocity components. The budget of the turbulent kinetic energy demonstrates the complex modification of the balance along both walls with a significant increase of production and dissipation downstream the reattachment point along the lower wall. This budget share a similar behavior at both walls, with and without separation. However the intensity of each term of the balance is much stronger at the lower wall. As reported by several authors, the production exhibits two peaks in the adverse pressure gradient region. The single peak observed before the summit of the bump is transported away from the wall and a second peak is created with the growth of a new boundary layer.

\section*{Acknowledgments}

Half of the simulations have been performed on IBM SP4 at IDRIS (CNRS computing facility, France) and the other half has been simulated with 128 processors of IBM SP5 at CRIHAN (Centre de Ressource en Informatique de Haute Normandie, France). This work was performed under the WALLTURB project. WALLTURB (A European synergy for the assessment of wall turbulence) is funded by the CEC under the $6^{th}$ framework program (CONTRACT \# AST4-CT-2005-516008).  We thank Dr. C. Braud and Prof. M. Stanislas for fruitful discussions and for their valuable comments on the manuscript.

\appendix

\section*{Appendix A: Mapping of coordinates}
\label{appendix}

In this appendix are given the transformed operators of the Navier-Stokes equations. For ease of writing, we define a new parameter:
\begin{equation}
\delta=\frac{\partial \eta}{\partial x} \;\; \frac{1}{\eta - L}
\end{equation}
The transformed gradient and Laplacian operator terms in equation (\ref{gradop}) write respectively:
\begin{equation} 
\vec{\nabla}_{\eta}=\left( \frac{\partial}{\partial x} \:,\: \frac{1-\gamma}{L} \frac{\partial}{\partial y}  \:,\: 
\frac{\partial}{\partial z} \right) \:\:,\:\: \vec{G}_{\eta} = \left( \delta (1-y) \: \frac{\partial}{\partial y} ,0,0 \right)
\end{equation}
\begin{equation} 
 \Delta_{\eta} = \frac{\partial^2}{\partial x^2} + \left(\frac{1-\gamma}{L} \right)^2  \frac{\partial^2}{\partial y^2} + \frac{\partial^2}{\partial z^2} \:\: ,
\end{equation}
\begin{equation}
L_{\eta} = \left( \frac{\partial \delta}{\partial x}-\delta^2 \right) \left( 1-y \right) \frac{\partial }{\partial y} + 2 \delta   \left( 1-y \right) \frac{\partial^2 }{\partial x \partial y} + \left( \delta   \left( 1-y \right) \right)^2 \frac{\partial^2 }{\partial y^2},
\end{equation}
The normal vector along the bump used to obtain the Neumann boundary condition for the intermediate pressure (\ref{neumnavier}) is:
\begin{equation}
\vec{\bar{n}} = \frac{1}{\sqrt{1+(\frac{\partial \eta}{\partial x})^2}} ( \vec{n} + \vec{n}_{\eta}) \:\: \mbox{with} \:\:
\vec{n}=(0,1) \:\: \mbox{and} \:\: \vec{n}_{\eta}=(-\frac{\partial \eta}{\partial x}, 0).
\end{equation} 
The second term of the r.h.s. of the Poisson equation for the pressure in equation (\ref{press}) writes:
\begin{eqnarray} 
 J (u ,v,w) =&&  2 \left[
\frac{1-\gamma}{L} \frac{\partial v}{\partial y} \left( \frac{\partial u}{\partial x} +
 \delta (1-y)  \frac{\partial u}{\partial y} \right) - 
 \frac{1-\gamma}{L} \frac{\partial u}{\partial y} \left( \frac{\partial v}{\partial x} +
 \delta (1-y) \frac{\partial v}{\partial y} \right) 
\nonumber \right.\\
&&
 - \frac{\partial u}{\partial z} \left( \frac{\partial w}{\partial x} +
 \delta (1-y) \frac{\partial w}{\partial y} \right)+
\frac{\partial w}{\partial z} \left(  \frac{\partial u}{\partial x}+
\delta (1-y)  \frac{\partial u}{\partial y} \right) 
\nonumber \\
&& \left.
- \frac{1-\gamma}{L} \frac{\partial w}{\partial y}
\frac{\partial v}{\partial z} + \frac{1-\gamma}{L}
\frac{\partial v}{\partial y}  \frac{\partial w}{\partial z} 
\right]. 
\end{eqnarray}  

\section*{Appendix B: Budget of the Reynolds Stress equations}

The equation of the Reynolds stress $\langle u_i^\prime u_j^\prime \rangle$ is obtained by multiplying the momentum equation for fluctuating velocity $u^\prime_i$ by $u^\prime_j$, multiplying the equation for $u^\prime_j$ by $u^\prime_i$ and adding. It reads

\begin{equation}
\frac{\partial \langle u_i^\prime u_j^\prime \rangle}{\partial t}   + \langle u_k \rangle \frac{\partial \langle u_i^\prime u_j^\prime \rangle}{\partial x_k}  = P_{ij} + T_{ij} + D_{ij} + D_{\rho,ij} + \Phi_{ij}- \epsilon_{ij}
\end{equation}
where the left-hand term is the advection and the right-hand terms are defined as follows:
\begin{eqnarray}
 \mbox{Production:} \hspace{2.6cm} P_{ij} &=& -\langle u_j^\prime u_k^\prime \rangle \frac{\partial \langle u_i \rangle}{\partial x_k} 
                                              -\langle u_i^\prime u_k^\prime \rangle \frac{\partial \langle u_j \rangle}{\partial x_k}, \\
 \mbox{Turbulent Transport: \hspace{1cm} }  T_{ij} &=&  - \frac{\partial \langle  u_i^\prime u_j^\prime u_k^\prime \rangle}{\partial x_k}, \\
 \mbox{Viscous Diffusion: \hspace{1.6cm}}  D_{ij} &=&  \nu \frac{\partial^2 \langle  u_i^\prime u_j^\prime \rangle}{\partial x_k \partial x_k}, \\
 \mbox{Pressure Diffusion: \hspace{1.2cm}}  D_{\rho,ij} &=& - \frac{1}{\rho} \left( \frac{\partial \langle  u_j^\prime p^\prime \rangle}{\partial x_i} +
                                                                    \frac{\partial \langle  u_i^\prime p^\prime \rangle}{\partial x_j} \right) , \\
 \mbox{Pressure Strain: \hspace{1.9cm}}  \Phi_{ij} &=& \left\langle \frac{p^\prime}{\rho} \left( \frac{\partial u_i^\prime}{\partial x_j} + \frac{\partial u_j^\prime}{\partial x_i} \right) \right\rangle, \\
 \mbox{Dissipation: \hspace{2.7cm}} \epsilon_{ij} &=& 2 \nu \left\langle  \frac{\partial u_i^\prime}{\partial x_k}  \frac{\partial u_j^\prime}{\partial x_k} \right\rangle.
\end{eqnarray}

The budget of the turbulent kinetic energy equation $k=\frac{1}{2} \langle u_i^\prime u_i^\prime \rangle$ is obtained by contracting $i$ and $j$ in the above equations.



\end{document}